%% file: main.tex
\documentclass[twocolumn]{aastex7}

\newcommand\unit{\mathrm}

\usepackage{graphicx}

\usepackage{amsmath}
\usepackage{url}
\usepackage{subfigure}
\usepackage{xspace}
\usepackage{enumitem}
\usepackage{bm}
\graphicspath{{figures/}{./}}
\makeatletter
\g@addto@macro\normalsize{%
  \setlength\abovedisplayskip{6pt plus 3pt minus 3pt}%
  \setlength\belowdisplayskip{6pt plus 3pt minus 3pt}%
  \setlength\abovedisplayshortskip{3pt plus 2pt minus 2pt}%
  \setlength\belowdisplayshortskip{3pt plus 2pt minus 2pt}%
}
\makeatother

\shortauthors{Tregoning \& Andrews}
\newcommand{\gaia}{{Gaia}\xspace}
\newcommand{\M}{{M$_{\odot}$}}

\begin{document}
\title{The Impact of Flyby Interactions on the Rotation of Open Clusters}
\correspondingauthor{Kyle R. Tregoning}
\email{kyle.tregoning.astro@gmail.com}

\author[0009-0009-6352-4964]{Kyle R. Tregoning}
\affiliation{Department of Physics, University of Florida, 2001 Museum Rd, Gainesville, FL 32611, USA}
\email{kyle.tregoning@ufl.edu}

\author[0000-0001-5261-3923]{Jeff J. Andrews}
\affiliation{Department of Physics, University of Florida, 2001 Museum Rd, Gainesville, FL 32611, USA}
\affiliation{Institute for Fundamental Theory, 2001 Museum Rd, Gainesville, FL 32611, USA}
\email{jeffrey.andrews@ufl.edu}

\begin{abstract}

The \gaia astrometric mission has provided the first evidence for the bulk rotation of open clusters. While this rotation may arise from residual spin inherited from a cluster's parent molecular cloud, it is unclear whether this explanation always suffices, as angular momentum is carried away by escaping stars on the timescale of two-body relaxation. In particular, the recent observance of significant rotation ($v_{\rm rot} = 0.81 \pm 0.06 ~\unit{km ~s^{-1}}$) in the dynamically old ($\approx 4 ~\unit{Gyr}$) open cluster M67 motivates a study of another potential source of spin for Solar neighborhood clusters: tidal torquing from gravitational encounters with passing giant molecular clouds (GMCs). We derive an analytic estimate for the change in angular momentum of a cluster due to an impulsive interaction with a passing GMC, and verify this expectation by performing $N$-body simulations over a range of encounter parameters using \texttt{REBOUND}. For highly energetic encounters, significant angular momentum can be transferred, but the stars that receive the most angular momentum are ejected, leaving the cluster remnant with negligible net rotation. Even when considering a random walk of many GMC--cluster encounters, the integrated effects of these interactions cannot explain observations of dynamically old rotating open clusters. However, we find that within $\sim 5$ Myr post-encounter, some disrupted clusters could be confused as rotating clusters while their stars are still dispersing. Using the local density of GMCs, we estimate that up to $\approx 1\%$ of open clusters in the Milky Way may carry this faux-rotation signal.

\end{abstract}

\section{Introduction} \label{sec:introduction}
\input{introduction}

\section{Methods}\label{sec:methods}
\input{methods}

\section{Results}\label{sec:results}
\input{results}

\section{Discussion}\label{sec:discussion}
\input{discussion}

\section{Conclusion}\label{sec:conclusion}
\input{conclusions}

\section*{Acknowledgment}

We thank Enrico Vesperini, Anna Lisa Varri, Carl Rodriguez, Ilya Mandel, Fred Rasio, and Brian Cook for useful discussions.

This research was supported by the Institute for Fundamental Theory scholarship through the Department of Physics at the University of Florida. The authors acknowledge UFIT Research Computing for providing computation resources and support that have contributed to the research results reported in this publication (\url{https://it.ufl.edu/rc}).

This work has made use of data from the European Space Agency (ESA) mission \gaia (\url{https://www.cosmos.esa.int/gaia}), processed by the \gaia Data Processing and Analysis Consortium (DPAC, \url{https://www.cosmos.esa.int/web/gaia/dpac/consortium}). Funding for the DPAC has been provided by national institutions, in particular the institutions participating in the \gaia Multilateral Agreement.

K.T.\ used ChatGPT (OpenAI) to assist in developing the simulation code, and Claude (Anthropic) to assist with intermediate steps in the analytic derivations, plotting code, and editing of the manuscript text. All AI-assisted content was reviewed and verified by K.T., and the authors take full responsibility for the work.

\software{\texttt{AstroPy} \citep{Astropy}, \texttt{Matplotlib} \citep{matplotlib}, \texttt{NumPy} \citep{numpy}, \texttt{REBOUND} \citep{rebound}, \texttt{REBOUNDx} \citep{reboundx}, \texttt{SciPy} \citep{SciPy}}

\appendix
\twocolumngrid
\section{Derivation of the Change in Angular Momentum}\label{app:appendix}
\input{appendix}

\bibliography{main}{}
\bibliographystyle{aasjournal}

\end{document}

%% file: introduction.tex
Coeval clusters of stars have long been utilized as fundamental laboratories for astrophysics. In particular, dynamical signatures of these clusters inform our knowledge across numerous domains; the rotation rates of constituent stars constrain gyrochronology \citep[e.g.,][]{Curtis_2020, Fritzewski_2021, Bouma_2023}, their dissolution constrains the tidal field they reside within \citep[e.g.,][]{Sanderson_2015, Meng_2022} and the inhomogeneities within it \citep[e.g.,][]{Gieles_2006, Bonaca_2019, Miller_2025}, samples of globular clusters reveal the assembly history of our Milky Way \citep[e.g.,][]{Kruijssen_2019, Massari_2019}, kinematics of young clusters reveal substructure in star forming environments \citep[e.g.,][]{Grosschedl_2018, Kuhn_2019}, and their internal kinematics inform us of expectations for gravitational wave detections \citep[e.g.,][]{Rodriguez_2016}.

One such dynamical signature, bulk rotation of the constituent stars about a central axis, has been studied extensively in clusters via numerical simulations. Orbit-averaged solutions to the 2D Fokker-Planck equations show that initially rotating clusters experience an accelerated mass loss and subsequent core-collapse \citep{Einsel_1999}, and that rotating clusters evolve more rapidly \citep{Kim_2002}. Simulations of clusters in a tidal field reveal the preferential escape of stars on prograde orbits relative to the clusters' orbital velocities \citep{Keenan_1975, Baumgardt_2003, Tiongco_2016}, the long-term evolution of bulk rotation towards an approximately solid-body limit equal to one-half the clusters' orbital frequencies \citep{Tiongco_2016, Tiongco_2018}, and the two-body driven degradation of primordial angular momentum and spin through the clusters' lifetimes \citep{Jerabkova_2021, Livernois_2022, Bianchini_2026} which occurs on the order of the half-mass relaxation time \citep{Bissekenov_2025}.

These simulations, by and large, have been performed with globular clusters (GCs) in mind; rotational detections in GCs \citep[e.g.,][]{Boberg_2017, Lanzoni_2018, Sollima_2019} are more robust than those observed in open clusters (OCs). Their high star counts and advantageous location out of the Galactic plane allow for significant ($>3\sigma$) rotation detections \citep[e.g.,][]{Bianchini_2018, Kamann_2018}. 

The ever-increasing astrometric precision of \gaia \citep{Gaia_DR3}, however, has finally shed light onto rotation in OCs by revealing correlated proper motions and/or radial velocities. \cite{Guilherme-Garcia_2023} detect rotation in eight OCs, with an additional nine candidates. \cite{Hao_2022, Hao_2024} report $\geq 3\sigma$ rotation detections in four nearby open clusters: Alpha Persei, the Pleiades, Praseape, and the Hyades. \cite{Jadhav_2024} report rotation signatures for 10 OCs (with 16 additional candidates), albeit at a lower precision (typically $<2\sigma$). \cite{Dalessandro_2026} recently published the largest rotating OC sample to date; they find significant rotation across a range of cluster ages, that the rotation signal weakens as a function of age, and that older OCs may be slightly more likely to display prograde rotation, though the overall alignment between the orbital and internal angular momentum vectors is still weak. These results are in broad agreement with the expectation from theory that older clusters should slowly lose their spin on a timescale comparable to the two-body relaxation time \citep[e.g.,][]{Bissekenov_2025}. 

The source of cluster rotation is often attributed to the primordial spin of their parent molecular clouds \citep[e.g.,][]{Henault-Brunet_2012, Kamann_2018, Bianchini_2018}. Velocity gradients indicative of net rotation in giant molecular clouds have been characterized both by observation \citep[e.g.,][]{Imara_2011, Braine_2018} and hydrodynamic simulations \citep[e.g.,][]{Mapelli_2017, Lahen_2020}. One such rotation detection in particular, however, challenges this paradigm; \cite{Sharma_2025} report a mean rotational velocity in the old open cluster M67 \citep[$\approx 4 ~\unit{Gyr}$,][]{Reyes_2024} of $0.81 \pm 0.06~\unit{km ~s^{-1}}$ within its Jacobi radius. \cite{Hurley_2005} find that M67 is dynamically old, having undergone $\approx 13$ relaxation times and having lost $\approx 90 \%$ of its initial mass. \cite{Tiongco_2016, Tiongco_2018} find that clusters orbiting in a tidal field born with primordial rotation evolve towards solid body rotation with an asymptotic angular velocity of $\Omega / 2$, where $\Omega$ is the angular velocity of the cluster about the host galaxy. Given the size and orbital velocity of M67, this corresponds to a rotational velocity of order $\sim 0.1 ~\unit{km ~s^{-1}}$, making the observed rotational velocity far larger than expected. M67 is not the only cluster challenging theoretical expectations for spin down; \cite{Dalessandro_2026} find at least $10$ clusters classified as significant rotators with ages greater than $1 ~\unit{Gyr}$. Note that M67 does not appear in this rotation sample, but \cite{Dalessandro_2026} make cuts to ignore clusters with prominent extra-tidal features, which M67 has \citep{Gao_2020, Sharma_2025}. Additionally, both \cite{Guilherme-Garcia_2023} and \cite{Jadhav_2024} classify Ruprecht 147 \citep[$\approx 3 ~\unit{Gyr}$ old;][]{Curtis_2013} as a rotating OC. We choose to highlight M67 specifically in this work due to its large reported rotational velocity, its old age, and the extensive studies of its dynamical history in the literature \citep[e.g.,][]{Hurley_2005, Carrera_2019}. 

Motivated both by these observations and the pending \gaia DR4 release, we present a theoretical framework, along with numerical validation, to describe another potentially significant source of cluster rotation: tidal shocking from flyby interactions with massive galactic over-densities. We tailor our analysis for OCs, and thus consider interactions with giant molecular clouds (GMCs) in this work, although our methodology could be abstracted to any star cluster interacting with a massive over-density. In Section~\ref{sec:methods}, we derive an analytic approximation for the angular momentum donated to stars via a GMC--cluster encounter and describe the initial conditions and methodology behind our $N$-body simulations of GMC--cluster encounters. We present the resultant changes in energy, mass, angular momentum, and rotational velocity of our simulated clusters in Section~\ref{sec:results}. We explore the limitations of our analytic model in Section~\ref{sec:discussion} and speculate on the existence of observations of dynamically relaxed OCs as well. Finally, we provide our concluding thoughts in Section~\ref{sec:conclusion}.

%% file: methods.tex
\subsection{Spin-Up via Individual Flyby Interactions}\label{subsec: spin up}

\begin{figure*}
    \centering
    \includegraphics[width=1\textwidth]{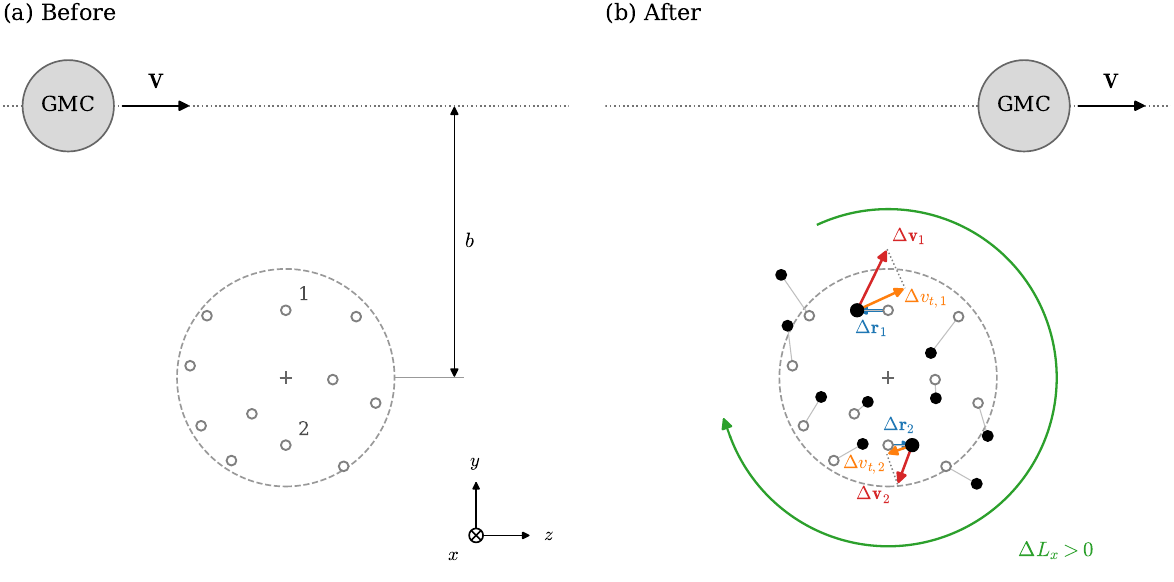}
    \caption{A schematic depicting the change in angular momentum of an initially spherical star cluster due to a flyby encounter with a passing GMC. We show the cluster both before (gray points) and after the encounter (black points), and specifically highlight the relative changes in position and velocity for two stars. Star 1, which is initially closer to the GMC than the cluster's center of mass, has an impact parameter less than $b$, and experiences a relative displacement in the $- \hat{z}$ direction and a relative velocity kick in the $\hat{y}$ and $\hat{z}$ directions when compared with star 2, which has an impact parameter greater than $b$. The asymmetry in both the position and velocity kicks lead to changes in angular momentum of the same order, which we refer to as $\Delta \mathbf{L}^{(r)}$ and $\Delta \mathbf{L}^{(v)}$, respectively. }
    \label{fig: schematic}
\end{figure*}

Consider a spherical star cluster of mass $M_c$ described by a Plummer density profile \citep{Plummer_1911} with half-mass radius $r_h$, at rest at the origin with an oncoming GMC with mass $M_{\rm GMC}$, impact parameter at infinity $b$, and relative velocity at infinity $V$. To derive the expected change in internal cluster angular momentum for a given GMC--cluster interaction, we adopt the following simplifying assumptions to derive an analytic, closed-form expression:
\begin{enumerate}[label=(\arabic*)]
    \item The impulse approximation is valid, i.e., $t_{\rm enc} \ll t_{\rm dyn}$, where the encounter time is defined as $t_{\rm enc} \equiv b/V$. We determine the dynamical time of the cluster from \cite{Spitzer_1987} as
    \begin{equation}\label{eq: t_dyn}
        t_{\rm dyn} \sim 2 \times 10^{-2} ~\unit{Myr} \left( \frac{M_c}{10^6 ~\unit{M_{\odot}}} \right)^{-1/2} \left( \frac{r_{\rm vir}}{1 ~\unit{pc}} \right)^{3/2},
    \end{equation}
    where virial equilibrium is assumed. For a Plummer profile, the virial radius $r_{\rm vir}$ is equated to the half-mass radius by $r_{\rm vir}=1.3 r_h$ \citep[e.g.,][]{Heggie_2003}. We thus ignore any orbital motion of stars during the encounter. 

    \item The encounter is weak, such that $b \gg b_{90}$, where $b_{90} = G (M_c + M_{\rm GMC}) / V^2 \approx G M_{\rm GMC}/V^2$, and $M_{\rm GMC} \gg M_c$. 
    
    \item The impact parameter is much larger than the size of the system, i.e., $b \gg r_h$, such that the GMC can be treated as a point mass.
\end{enumerate}

In this work, we focus solely on a cluster of mass $M_c = 1000 ~\unit{M_{\odot}}$, impact parameters $b < 100 ~\unit{pc}$, and a relative velocity at infinity of $V = 20 ~\unit{km ~s^{-1}}$. At these parameters, note that $t_{\rm enc} < 4.9 ~\unit{Myr}$ and $t_{\rm dyn} \approx 4.9 ~\unit{Myr}$, computed from Equation~\ref{eq: t_dyn}. Though assumption (1) is not always strictly satisfied at every simulated impact parameter, $N$-body simulations show that the impulse approximation is a fairly reasonable approximation of the relative energy gain for $t_{\rm enc} \leq t_{\rm dyn}$ for a point mass perturber \citep[see the right panel of Figure 1 in][]{Martinez-Medina_2022}. Additionally, the stars that receive the largest kicks relative to the cluster center of mass are those at large radii, where the local dynamical time is larger. Thus, we consider the impulse approximation a reasonable assumption in computing the angular momentum change for all simulations explored hereinafter.

We orient the GMC's velocity along the $z$-axis, and place the GMC's orbit in the $yz$-plane, such that the GMC is at position $(0, b, 0)$ at closest approach. We assume the GMC travels on a straight-line trajectory, unaffected by the star cluster's gravity, which is a reasonable assumption for a hyperbolic trajectory with $M_{\rm GMC} \gg M_c$. A star in the cluster with initial position $\mathbf{r}_i = (x_i, y_i, z_i)$ relative to the cluster origin will experience both a velocity kick $\Delta \mathbf{v}_i$ relative to the cluster center of mass velocity, as well as a displacement relative to the center of mass $\Delta \mathbf{r}_i$ caused by the acceleration from the GMC during the encounter. The change in angular momentum of the cluster as a whole, relative to its center of mass, is then
\begin{equation}\label{eq: delta_L_def}
    \Delta \mathbf{L} = \sum_i m_i (\mathbf{r}_i +\Delta \mathbf{r}_i) \times \Delta \mathbf{v}_i = \Delta \mathbf{L}^{(v)} + \Delta \mathbf{L}^{(r)} 
\end{equation}
where $\Delta \mathbf{L}^{(v)} \equiv \sum_i m_i \mathbf{r}_i \times \Delta \mathbf{v}_i$ and $\Delta \mathbf{L}^{(r)} \equiv \sum_i m_i \Delta \mathbf{r}_i \times \Delta \mathbf{v}_i$. 

Starting from Equations (3.54a) and (3.54b) in \cite{Binney_Tremaine_2008}, which give the final velocity kick for an object in the center of mass frame from a hyperbolic encounter, we can approximate the change in angular momentum for the star cluster as 
\begin{equation}\label{eq: delta_Lx_int}
\begin{aligned}
    \Delta L_x^{\rm int} &= \Delta L_x^{(v)} + \Delta L_x^{(r)} \\
    &\approx 2V \int \frac{y}{1 + (b_{\rm eff}/b_{90})^2}\, \rho(r)\, d^3 r \\
    &\quad - 2 b_{90} \int \ln\!\left( \frac{b_{\rm eff}}{b} \right) \Delta v_y(x, y)\, \rho(r)\, d^3 r
\end{aligned}
\end{equation}
where $b_{\rm eff} \equiv \sqrt{x^2 + (b - y)^2}$, $\rho(r)$ is the Plummer density profile $\rho(r) = (3 M_c / 4\pi a^3 ) \left[ 1 + r^2 / a^2 \right]^{-5/2}$, and $a$ is the Plummer radius, related to the half-mass radius by $r_h = 1.305 a$. We derive Equations \ref{eq: delta_L_def} and \ref{eq: delta_Lx_int} in detail in Appendix \ref{app:appendix}, and note that the only surviving term of order $b_{90}^2$ or lower is the $x$-component of $\Delta \mathbf{L}$ under our chosen coordinate system for the encounter. 

Following assumptions (2) and (3), we can derive a closed form expression for the cluster's change in internal angular momentum. Writing   
\begin{equation}\label{eq: approxs}
    b_{\rm eff}^2 = b^2 - 2 b y + x^2 + y^2 \equiv b^2 (1 + \epsilon), ~\epsilon = - \frac{2y}{b} + \frac{x^2 + y^2}{b^2},
\end{equation}
where $\epsilon \ll 1$, and keeping all leading order terms following expansion about $b_{90} / b$ and $\epsilon$, reduces Equation~\ref{eq: delta_Lx_int} to 
\begin{equation}\label{eq: delta_Lx_approx}
    \Delta L_x^{\rm approx} \approx \frac{8 V b_{90}^2}{b^3} \int y^2 \rho(r) d^3r = \frac{8}{3} \frac{G^2 M_{\rm GMC}^2 M_c }{V^3 b^3 } \langle r^2 \rangle,
\end{equation}
where $\langle r^2 \rangle$ is the mean squared radius of stars in the cluster. Note that, at given encounter parameters $M_{\rm GMC}$, $b$, and $V$, the change in angular momentum of the cluster is proportional to $M_c \langle r^2 \rangle$, which is some factor multiplied by the cluster's moment of inertia. 

In Figure~\ref{fig: schematic}, we show a schematic of a flyby encounter between a GMC and a star cluster in our chosen coordinate system to demonstrate how the angular momentum of the cluster changes. We show the cluster both before (gray points) and after (black points) the interaction, and focus on the position and velocity displacements relative to the center of mass motion of two stars in particular; one that sits between the center of mass and the GMC (star 1) and another that sits below the center of mass (star 2). Relative to star 2, star 1 is pulled towards the $-z$ direction during the encounter, and receives a velocity kick in the $y$ and $z$ directions. Both the displacement and the velocity kick of star 1 relative to star 2 lead to changes in angular momentum of the same order in the $x$ direction. 

\begin{figure}
    \centering
    \includegraphics[width=1\columnwidth]{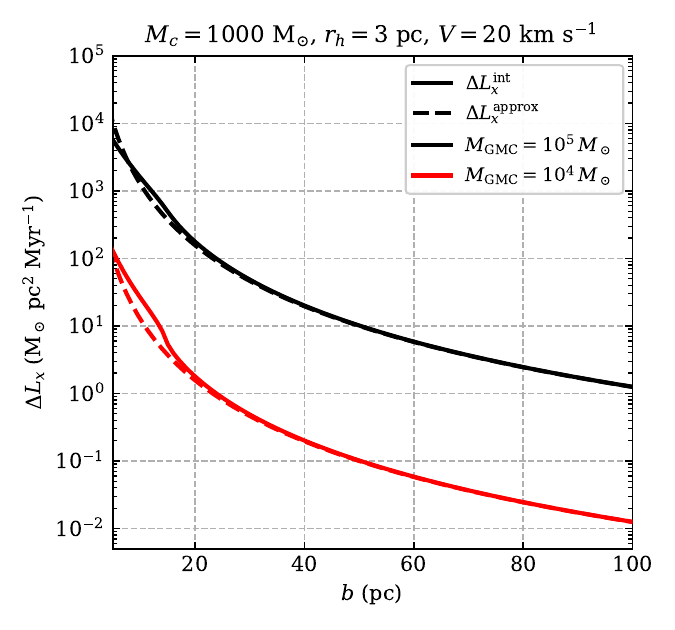}
    \caption{The expected change in angular momentum in the $x$ direction for an encounter in the $yz$-plane between a GMC and a spherical star cluster with mass $M_c = 1000 ~\unit{M_{\odot}}$, $r_h = 3 ~\unit{pc}$, and relative velocity at infinity $V = 20 ~\unit{km ~s^{-1}}$, as a function of impact parameter $b$. Solid lines show the numerical solution to the integral formulation of $\Delta L_x$ (Equation~\ref{eq: delta_Lx_int}), while dashed lines are computed from the weak-encounter approximation (Equation~\ref{eq: delta_Lx_approx}). We truncate the cluster at a tidal radius of $r_t = 14.75 ~\unit{pc}$. Here we show two different GMC masses; $M_{\rm GMC} = 10^4 ~\unit{M_{\odot}}$ (red) and $M_{\rm GMC} = 10^5 ~\unit{M_{\odot}}$ (black).}
    \label{fig: dLx}
\end{figure}

In Figure~\ref{fig: dLx}, we plot the change in angular momentum in the $x$ direction $\Delta L_x$ as a function of impact parameter $b$ for a cluster of mass $1000 ~\unit{M_{\odot}}$, half-mass radius $r_h = 3 ~\unit{pc}$, and relative encounter velocity $V = 20 ~\unit{km ~s^{-1}}$. Solid lines show the numerical solution to Equation~\ref{eq: delta_Lx_int}; for the radial portion of the integral, we integrate from $r=0$ to $r=r_t = 14.75 ~\unit{pc}$, which is the truncation radius used when generating initial conditions for our $N$-body models (see Section~\ref{subsec: sims}). Dashed lines refer to Equation~\ref{eq: delta_Lx_approx}, where we have solved for the mean squared radius as $\langle r^2 \rangle = M_c^{-1}\int_0^{r_t} r^2 \rho(r) d^3r $ using a Plummer density profile. The two formulations of $\Delta L_x$ agree well for $b \gtrsim 20 ~\unit{pc}$ at our chosen GMC masses of $10^4 ~\unit{M_{\odot}}$ and $10^5 ~\unit{M_{\odot}}$. 

\subsection{N-body Simulations}\label{subsec: sims}
To simulate encounters between clusters and GMCs, we utilize \texttt{REBOUND} \citep{rebound} to perform all $N$-body simulations, using the \texttt{IAS15} integrator \citep{reboundias15}; a non-symplectic, 15th order Gauss-Radau integrator with adaptive time-stepping. We store simulation outputs using the \texttt{Simulationarchive} \citep{reboundsa}. 

We initialize star clusters using \texttt{McLuster} \citep{mcluster}. Each star cluster is generated from a Plummer profile \citep{Plummer_1911} with a total mass $M_c = 1000 ~\unit{M_{\odot}}$ and half-mass radius $r_h = 3 ~\unit{pc}$, which corresponds to a Plummer radius $a_c = 2.3 ~\unit{pc}$. By default, these models are truncated at a tidal radius of $r_t = 14.75 ~\unit{pc}$. We generate stellar masses using the canonical Kroupa \citep{Kroupa_2001} initial mass function, with a minimum and maximum stellar mass of $0.08 ~\unit{M_{\odot}}$ and $100 ~\unit{M_{\odot}}$, respectively. We do not consider any primordial binaries or mass segregation, nor do we consider stellar evolution in this work. 

Throughout all simulations, we treat the GMC as a point mass. Galactic GMCs are found to be extended objects with sizes roughly proportional to $M_{\rm GMC}^{1/2}$ \citep[e.g.,][]{Solomon_1987, Lada_2020, Luo_2026}, such that a typical GMC of mass $10^5 ~\unit{M_{\odot}}$ is $\sim 30 ~\unit{pc}$ in extent \citep[see][]{Lada_2020}. For simulations with $b < 30 ~\unit{pc}$, the energy and angular momentum gain of the simulated cluster will be larger than it would have been had the GMC been ascribed a realistic size. Thus, we consider the results of our simulations to be an upper limit on the angular momentum gain of a cluster that undergoes an encounter with a given set of impact parameters, as exploring the effect of the GMC size on the imparted angular momentum is outside of the scope of this work. 

We carry out the GMC--cluster encounter in the same manner as \cite{Gieles_2006}. That is, we begin the simulation with the cluster centered at the origin, and the GMC placed at position $(0, b, -D)$ with velocity $(0, 0, V)$, where $D = 500 a_c$, corresponding to $\approx 1150 ~\unit{pc}$ given our chosen cluster initialization. Here, $b$ and $V$ are the impact parameter and relative velocity at infinity, respectively. The GMC is placed on a straight-line trajectory, such that $\mathbf{r}_{\rm GMC}(t) = (0, b, -D + V t) $. In all simulations, the GMC is at least $10$ times more massive than the cluster, making a straight-line path a reasonable approximation. Note that, while the GMC is on a fixed trajectory, the cluster's center of mass is allowed to respond to the GMC, such that the closest approach distance and relative velocity at impact are smaller and larger than $b$ and $V$, respectively. 

To avoid sudden jumps in the potential felt by simulated stars, we ramp up (and down) the mass of the GMC in a smooth fashion, based on the distance $r_{\rm GMC}$ from the GMC to the origin. The simulated GMC mass $M_{\rm GMC}^{\star}$ as a function of $r_{\rm GMC}$ is
\begin{equation}
M_{\rm GMC}^{\star}(r_{\rm GMC}) =
\begin{cases}
0, & r_{\rm GMC} \geq D,\\
\frac{M_{\rm GMC}}{2}[1-\cos(\pi x)],
& 0.9D < r_{\rm GMC} < D,\\
M_{\rm GMC}, & r_{\rm GMC} \leq 0.9D.
\end{cases}
\label{eq: piecewise}
\end{equation}
where $x \equiv 10 (D - r_{\rm GMC})/D$, and $M_{\rm GMC}$ is the peak mass of the GMC. We incorporate the simulated GMC as an external force in our simulations, not an additional particle, utilizing \texttt{REBOUNDx} \citep{reboundx} to do so. 

The total duration of each GMC--cluster simulation is three times the amount of time taken for the GMC to reach position $(0, b, 0)$, i.e., $T_{\rm total} = 3 T_{\rm enc}$, where $T_{\rm enc} \equiv D/V$ (not to be confused with the relevant encounter timescale $t_{\rm enc} \equiv b /V$). This gives stars with velocities greater than the escape velocity sufficient time to leave the cluster post-encounter. 

In this work, we present GMC--cluster simulations with
$M_{\rm GMC} = 10^4$ and $10^5 ~\unit{M_\odot}$, impact parameters $b = 5$, $10$, $20$, $40$, $60$, $80$, and $100 ~\unit{pc}$, and a constant relative velocity at infinity of $V = 20 ~\unit{km ~s^{-1}}$. For each of the possible $14$ combinations of $M_{\rm GMC}$, $b$, and $V$, we run $10$ realizations with different initial seeds for the cluster. At our chosen $V$, each simulation runs for a total of $T_{\rm total} = 168.6 ~\unit{Myr}$. For comparison, we also run $10$ simulations where a cluster with the same initial conditions evolves in isolation for the same time $T_{\rm total} = 168.6 ~\unit{Myr}$, which corresponds to $\approx 2$ initial half-mass relaxation times for a cluster with a Plummer density profile with mass $M_c = 1000 ~\unit{M_{\odot}}$ and $r_h = 3 ~\unit{pc}$.  

%% file: results.tex
\subsection{Energy Gain and Mass Loss from the Cluster}\label{subsec: energy and mass}

\begin{figure*}
    \centering
    \includegraphics[width=1\textwidth]{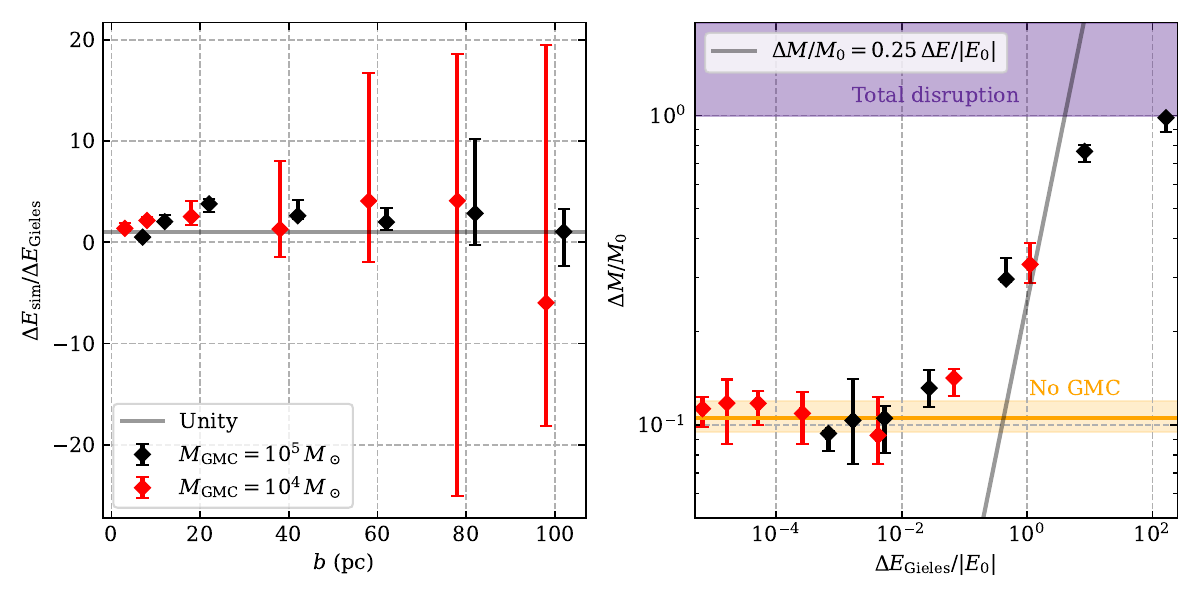}
    \caption{\textit{Left}: The change in internal energy of all stars for each simulated encounter divided by the expected change in internal energy predicted by \cite{Gieles_2006}, Equation (15), plotted against the impact parameter $b$. Each filled diamond represents the median value of $\Delta E_{\rm sim} / \Delta E_{\rm Gieles}$ over the $10$ simulations at that given $b$ and $M_{\rm GMC}$. The horizontal gray line represents perfect agreement between the simulations and analytic prediction. The error bars are representative of the 25th--75th percentile spread in $\Delta E_{\rm sim} / \Delta E_{\rm Gieles}$ over that given set of $10$ simulations. Note that we add an artificial horizontal offset of $\pm 2 ~\unit{pc}$ from the true impact parameter for visual clarity. \textit{Right}: The fractional mass loss of the cluster, where $\Delta M \equiv M_0 - M_f$, as a function of the predicted relative energy increase (where $E_0$ is the initial cluster total energy) due to the GMC--cluster encounter. As in the left panel, filled diamonds represent the median value, while error bars represent that 25th--75th percentile spread. The shaded purple region represents complete cluster dissolution. The gray line shows the scaling relation between fractional mass loss and fractional energy gain found by \cite{Gieles_2006}; note that our simulations cover a much larger region of $\Delta E/ E_0$ space than those covered in their work. For comparison, the orange line is the median fractional mass loss for simulations where the cluster evolves in isolation, while the orange band represents the 25th--75th percentile spread of these simulations. For weak encounters with $\Delta E / |E_0| < 10^{-2}$, the encounter is negligible in terms of the cluster mass loss, as the mass loss is fully consistent with two-body relaxation.}
    \label{fig: dE_dM}
\end{figure*}

In the left panel of Figure~\ref{fig: dE_dM} we compare the relative energy gain in each simulation with the analytic prediction for such a GMC--cluster encounter by \cite{Gieles_2006} as a function of impact parameter $b$. The quantity $\Delta E_{\rm sim}$ is taken as the difference in total energy of all stars in the simulation at the beginning and end, i.e., $E_f - E_0$, computed internally via \texttt{REBOUND}, where the center of mass motion has been subtracted off. The analytic prediction $\Delta E_{\rm Gieles}$ is computed from their Equation (15) of \cite{Gieles_2006} using the half-mass radius $r_h$ at the initial timestep, after converting our impact parameters, which are defined at infinity, to theirs, which are defined at closest approach between cluster and GMC. Each diamond represents the median of $10$ simulations at a given $b$ and $M_{\rm GMC}$, and the error bars represent the 25th--75th percentile spread, i.e., the inter-quartile range, of $\Delta E_{\rm sim} / \Delta E_{\rm Gieles}$ for those $10$ simulations. We see good agreement at low ($b < 30 ~\unit{pc}$) impact parameters (and thus, a large relative energy gain), which is the most interesting regime in terms of adding net angular momentum to stars in the cluster (see Figure~\ref{fig: dLx}). At larger impact parameters, while our simulated results still show consistency with the predictions of \cite{Gieles_2006}, the spread in the energy gain increases dramatically. This is expected; \cite{Gieles_2006} predict only the second order change in the energy, as the first order change in internal energy $\Delta E^{(1)} = \sum_i m_i \mathbf{v}_i \cdot \Delta \mathbf{v}_i$ sums to zero for a perfectly isotropic velocity distribution. For a finite $N$ system, $\Delta E^{(1)}$ becomes comparable to $\Delta E_{\rm Gieles}$ for less energetic interactions (i.e., high $b$, low $M_{\rm GMC}$). Nonetheless, the median and spread of all simulations explored in this work see energy increases that broadly agree with theoretical expectation, validating our continued employment of the impulse approximation. 

In the right panel of Figure~\ref{fig: dE_dM}, we show the fractional mass loss of the cluster, where $\Delta M \equiv M_{c,0} - M_{c,f}$, as a function of the predicted relative energy gain of the encounter, where $E_{0}$ is the total initial cluster energy in the center of mass frame, computed internally via \texttt{REBOUND}. $M_{c,0}$ and $M_{c,f}$ refer to the bound mass at the initial and final timestep. Boundedness at a given timestep is determined iteratively; we compute the relative energy of each star as $\epsilon_i = \tfrac{1}{2}|\mathbf{v}_i - \bar{\mathbf{v}}|^2 + \Phi_i (\mathbf{r}_i)$, where $\mathbf{v}_i$ is the velocity of the $i$-th star relative to the center of mass velocity of the current bound subset, and $\Phi_i (\mathbf{r}_i)$ is the potential of the current bound subset at position $\mathbf{r}_i$, computed via direct summation. Stars with $\epsilon > 0$ are removed from the bound subset, and the procedure is repeated until the bound membership has converged. Filled diamonds again represent the median of $10$ simulations at a given $b$ and $M_{\rm GMC}$, and error bars represent the 25th--75th percentile spread. The gray line represents the empirical fit found by \cite{Gieles_2006} between the energy gain and mass loss, i.e., $\Delta M / M_0 \approx 0.25 \Delta E / |E_0|$. 

Our simulations cover a large range in $\Delta E / |E_0|$, while the simulations explored by \cite{Gieles_2006} probe relative energy gains from $10^{-3}$ to $10$. Our simulations with $\Delta E_{\rm Gieles} / |E_0| \approx 1$ agree well with their empirical fit, but deviate at larger and smaller relative energy gains. We also show, in orange, both the median and fractional spread in the fractional mass loss for our simulations where the cluster evolved in isolation, to control for the effects of mass lost via two-body relaxation. We see that, for weak encounters with $\Delta E_{\rm Gieles} / |E_0| < 10^{-2}$, the encounter is negligible in terms of causing mass loss, as the entirety of the mass lost from the cluster can be explained via two-body processes. 

\subsection{Angular Momentum}\label{subsec: angular momentum}

\begin{figure}
    \centering
    \includegraphics[width=1\columnwidth]{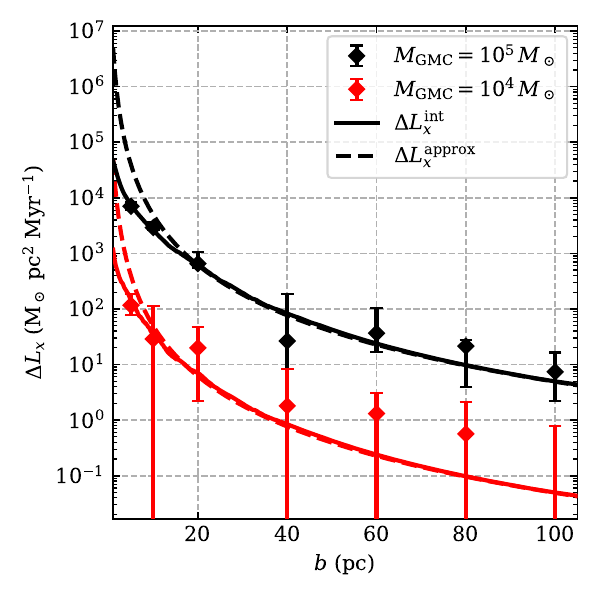}
    \caption{The change in angular momentum of all stars in the simulation, relative to the center of mass of the system. Filled diamonds represent the median of $10$ simulations at a given impact parameter $b$ and GMC mass $M_{\rm GMC}$, and error bars represent the 25th--75th percentile spread across those $10$ simulations. Dotted lines show the predicted change in angular momentum at a given $M_{\rm GMC}$ (Equation~\ref{eq: delta_Lx_approx}). For most simulation sets with $M_{\rm GMC} = 10^4 ~\unit{M_{\odot}}$, the spread of $\Delta L_x$ extends to negative values. Our simulations confirm that Equation~\ref{eq: delta_Lx_approx} provides a reasonable analytic approximation for $\Delta L_x$.}
    \label{fig: dLx_sim}
\end{figure}

In Figure~\ref{fig: dLx_sim}, we present a comparison of the change in angular momentum from our simulations with the analytic approximation for $\Delta L_x$ for weak encounters under the impulse approximation, i.e., Equations~\ref{eq: delta_Lx_int} (solid lines) and \ref{eq: delta_Lx_approx} (dashed lines). Filled diamonds represent the median change in internal angular momentum in the $x$-direction across a given subset of $10$ simulations, while error bars again represent the 25th--75th percentile spread. Both Equations~\ref{eq: delta_Lx_int} and \ref{eq: delta_Lx_approx} are computed using simulations snapshots two encounter times before the closest passage between cluster and GMC, i.e., $t_{\rm pre-enc} = T_{\rm enc} - 2 b /V$, where all positions and velocities have been shifted to the cluster's center of mass frame. We choose this timestep because the cluster has only been minimally perturbed by the GMC \citep[see Figure 6 in][]{Gieles_2006}, while also having relaxed from its initial conditions. We compute Equation~\ref{eq: delta_Lx_int} exactly; for each simulation Equation~\ref{eq: delta_Lx_int} is evaluated via direct summation over the stellar positions at $t_{\rm pre-enc}$ over a grid of $b$ values from $1$ to $105 ~\unit{pc}$. The solid red and black lines show the median of these curves for simulations with $M_{\rm GMC} = 10^4 ~\unit{M_{\odot}}$ and $M_{\rm GMC} = 10^5 ~\unit{M_{\odot}}$, respectively. To compute Equation~\ref{eq: delta_Lx_approx}, we evaluate $\langle r^2 \rangle$ relative to the center of mass at $t_{\rm pre-enc}$ for every simulation explored in this work, including the control set with no GMC. We use the median $\langle r^2 \rangle$ at $t_{\rm pre-enc}$ across all simulations to evaluate Equation~\ref{eq: delta_Lx_approx}, as taking $\langle r^2 \rangle$ at $t=0$ ignores the pre-encounter radial evolution of the cluster, and would under-predict $\Delta L_x$. Note that the angular momenta of all stars relative to the center of mass are considered here, regardless of whether or not the stars are unbound to the cluster by the encounter or by two-body processes.

For all encounters tested, we see good agreement between simulation and theory. At impact parameters $b > 40~\unit{pc}$, the change in angular momentum for clusters that have encounters with $10^4 ~\unit{M_{\odot}}$ GMCs are dominated by stochastic variability due to the finite number of stars. The first order contribution to the change in angular momentum $\Delta L_{x}^{(1)}$, which vanishes under perfect symmetry, is comparable to the signal $\Delta L_x$ in this regime. Since $\Delta L_x \propto M_{\rm GMC}^2$, encounters with $M_{\rm GMC} = 10^5 ~\unit{M_{\odot}}$ produce a signal two orders of magnitude larger than those with $M_{\rm GMC} = 10^4 ~\unit{M_{\odot}}$, and thus, the signal-to-noise ratio is adequate out to values of $b = 100 ~\unit{pc}$. For very close encounters, i.e., $b=5$, Equation~\ref{eq: delta_Lx_approx} over-predicts the angular momentum. At low impact parameters, the size of the GMC also becomes important; we expect our simulations and analytic prescription to provide an upper limit on $\Delta L_x$ in this regime, as a point mass approximation overestimates the energy gain in such an encounter \citep[see Figure 7 in][]{Gieles_2006}. 

While our results show that the angular momentum change of all stars due to a GMC--cluster encounter is predictable analytically, we aim to explore the rotational properties of the \textit{bound} subset of stars that remain in the cluster post-encounter. We consider a parameter more observationally obtainable than angular momentum, and thus define the rotational velocity at some epoch $v_{\rm rot}$ to be the mean of the tangential velocity of all bound stars about the cluster's principal axis of rotation ($v_{\phi}$), defined by the angular momentum direction $\hat{\mathbf{L}}$ at that epoch. This follows from the approach used to detect angular momentum in M67 by \cite{Sharma_2025}. 

\begin{figure*}
    \centering
    \includegraphics[width=1\textwidth]{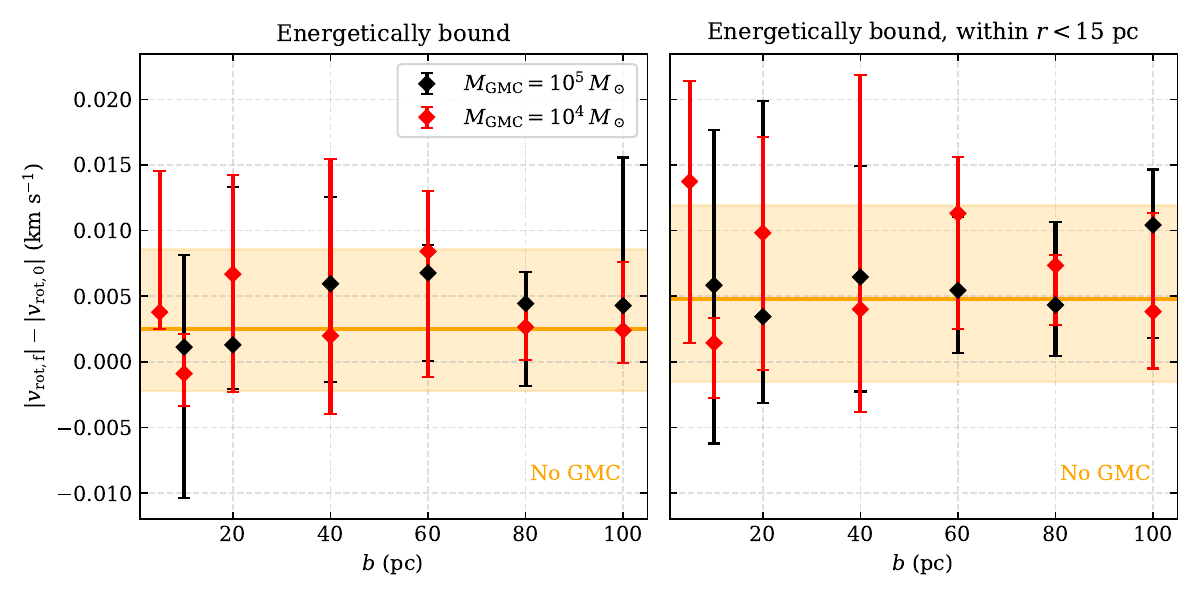}
    \caption{\textit{Left}: The magnitude of the final rotational velocity minus the magnitude of the initial rotational velocity, for the bound subset of stars at the simulation's beginning and end. Filled diamonds represent the median of up to $10$ simulations at a given impact parameter $b$ and GMC mass $M_{\rm GMC}$, and error bars represent the 25th--75th percentile spread across those simulations. The orange line and band represents the median and 25th--75th percentile spread for our control simulations where the cluster evolves in isolation. We include only simulations with $\geq 50$ bound members at simulation end. \textit{Right}: The same as the left panel, but only including stars with distances of $r < 15 ~\unit{pc}$ from the density center of the bound subset of stars. Across all simulated impact parameters and GMC masses, we find no appreciable ``spin-up" due to GMC--cluster encounters.}
    \label{fig: vrot}
\end{figure*}

In Figure~\ref{fig: vrot}, we show the change in the magnitude of the rotational velocity $v_{\rm rot}$ between the first and last snapshot for all simulations with $\geq 50$ bound members at simulation end. Again, we plot the median and 25th--75th percentile spread for a given simulation set.  We determine boundedness in the same manner as in Section~\ref{subsec: energy and mass}. At both the initial and final epoch, we determine the cluster density center using $k=32$ nearest neighbors from all bound stars. Positions are measured relative to this density center, and velocities are measured relative to the center of mass velocity of the bound subset of stars. In the left panel of Figure~\ref{fig: vrot}, we include all stars energetically bound to the system when determining $v_{\rm rot}$, while in the right panel, we only consider energetically bound stars within $15 ~\unit{pc}$ of the cluster density center. We also show, as an orange line and band, the same difference in magnitude of initial and final rotational velocity for simulations with no GMC--cluster encounter. The vast majority of clusters undergoing GMC encounters do not gain appreciable spin in their bound subset when compared with clusters subject only to two-body relaxation, implying such encounters are not a suitable mechanism to ``spin-up" clusters on a large scale. No single simulation experiences a gain in rotational velocity of $> 0.1 ~\unit{km ~s^{-1}}$, indicating that even the most optimal encounters do not appreciably cause an increase in cluster rotation after the stars unbound by the encounter have had sufficient time to leave the system. 

\subsection{Apparent Spin-Up of a Cluster}
\begin{figure*}
    \centering
    \includegraphics[width=1\textwidth]{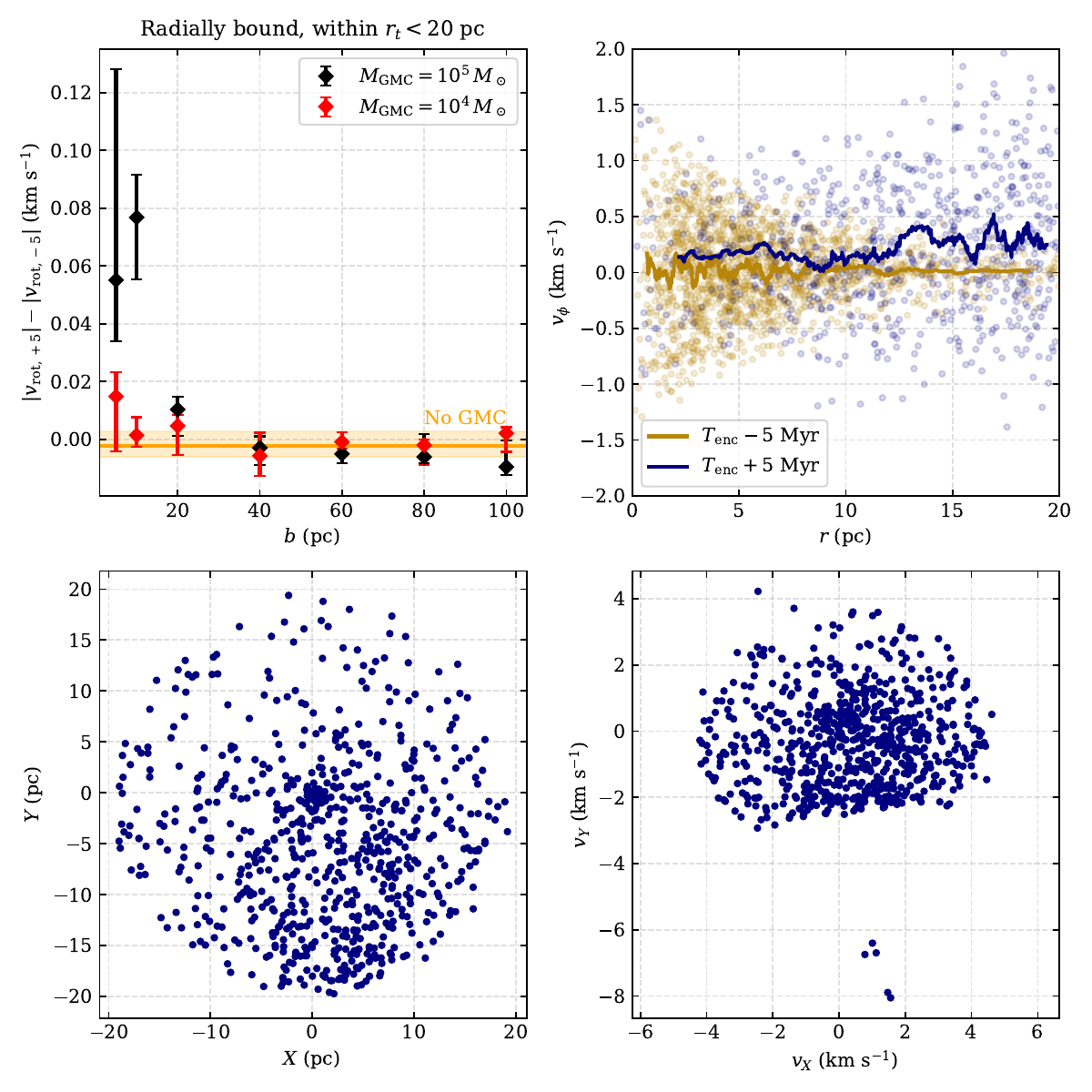}
    \caption{\textit{Top Left}: The same as Figure~\ref{fig: vrot}, but $v_{\rm rot}$ is defined $5 ~\unit{Myr}$ before and after the GMC passage, and bound stars are defined by a radial criterion of $r < r_t = 20 ~\unit{pc}$ as opposed to being energetically defined. At low impact parameters, some clusters gain appreciable rotational velocity compared with their rotational state immediately before the encounter. \textit{Top Right}: The velocity about the rotation axis, in one example simulation with $b = 5 ~\unit{pc}$ and $M_{\rm GMC} = 10^5 ~\unit{M_{\odot}}$, of all stars within $20 ~\unit{pc}$ of the cluster center, $5 ~\unit{Myr}$ before (gold) and after (navy) the encounter, plotted against the distance to the cluster density center. Solid lines show the rolling mean across $50$ points for each of the two epochs. Stars beyond $10 ~\unit{pc}$ post-encounter show net-positive values of $v_{\phi}$. \textit{Bottom Left}: The $X$ and $Y$ positions of stars in our example simulation within $20 ~\unit{pc}$ of the cluster center $5 ~\unit{Myr}$ post-encounter. \textit{Bottom Right}: The velocity in the $X$ and $Y$ directions of these same stars. At this epoch, the cluster still appears quasi-spherical in two dimensions of position and velocity space.}
    \label{fig: vrot_tidal_example}
\end{figure*}

While there is no considerable ``spin-up" of the cluster at simulation end, the rotational properties of the cluster a short time after the encounter are still of interest. In the top left panel of Figure~\ref{fig: vrot_tidal_example}, we instead compute the rotational velocity $5 ~\unit{Myr}$ before ($v_{\rm rot,-5}$) and after ($v_{\rm rot,+5}$) the time of encounter $T_{\rm enc}$ for each simulated cluster. The orange line and band represent the median and 25th--75th percentile spread of the control simulations, with rotational velocities measured $5 ~\unit{Myr}$ before and after the time at which a GMC would have reached closest approach. Throughout Figure~\ref{fig: vrot_tidal_example}, we define the bound subset of stars differently than in Figures~\ref{fig: dE_dM} and \ref{fig: vrot}. Here, we find the density center amongst all stars, bound or unbound energetically, using $k=32$ nearest neighbors, and define a tidal radius of $r_t = 20 ~\unit{pc}$ \cite[chosen as a comparative value with the Jacobi radius of M67 as reported by][]{Hunt_2024} relative to that density center; all stars within $r_t$ of the density center are considered radially bound. Positions are measured relative to this density center, and velocities are defined relative to the center of mass velocity of this radially bound subset in order to compute $\hat{\mathbf{L}}$ and $v_{\rm rot}$ at each given epoch. We consider only simulations with $\geq 50$ radially bound members. This approach is akin to how observers define bound and unbound members of a star cluster \citep[e.g.,][]{Hunt_2024} as finding the relative energy of individual stars requires accurate 6D phase space information, which is often very difficult to obtain. We see, in the top left panel of Figure~\ref{fig: vrot_tidal_example}, that when looking closely before and after the encounter and defining boundedness purely by a radial criterion, some simulations with high relative energy gains (i.e., smaller impact parameters) can produce appreciable rotation signatures when compared with the control simulations where no GMC is present. 

The other three panels in Figure~\ref{fig: vrot_tidal_example} highlight one example simulation of a cluster that has significant rotational velocity within $20 ~\unit{pc}$ of the density center $5 ~\unit{Myr}$ post-encounter. This particular simulation featured an encounter with impact parameter $b = 5 ~\unit{pc}$ and GMC mass $M_{\rm GMC} = 10^5 ~\unit{M_{\odot}}$. The top right panel shows the velocity about the rotation axis $v_{\phi}$ for each star, plotted against the 3D distance $r$ to the cluster density center. We show $v_{\phi}$ at $5 ~\unit{Myr}$ before (gold points) and after (navy points) the encounter, and display a rolling average as a solid line over $50$ points for each of the two epochs. Shortly after the GMC--cluster encounter, stars in the cluster outskirts beyond $10 ~\unit{pc}$ are characterized by a net positive value of $v_{\phi}$; the rolling mean of $v_{\phi}$ post-encounter reaches $\approx 0.5 ~\unit{km ~s^{-1}}$, within detectability limits for nearby OCs. The bottom two panels show the cluster in position and velocity space in the $xy$-plane at $5 ~\unit{Myr}$ after the encounter. This particular cluster still appears quasi-spherical in two dimensions in both position and velocity space despite the encounter, though the cluster develops a more elongated morphology in the $z$ direction due to the GMC passage. A cluster like this one, observed on the sky shortly after an energetic encounter, may appear as a spherical cluster with a detectable rotation signature, though much of this signature is produced by stars that have gained enough energy post-encounter to escape the system after a few crossing times. 

%% file: discussion.tex
The results of our simulations confirm the theoretical framework outlined to describe the amount of angular momentum donated from a passing GMC onto a spherical star cluster in the impulsive regime. However, our results show that this angular momentum does not appreciably ``spin-up" the cluster after stars energetically unbound from the system have been given ample time to leave it, as they carry away a significant fraction of the encounter-induced angular momentum. Naively, this makes sense, as the energy induced by such an encounter is proportional to $b^{-4}$ \citep[Equation 15 in][]{Gieles_2006} while the angular momentum gain from Equation~\ref{eq: delta_Lx_approx} is proportional to $b^{-3}$. There are two competing levers at a given relative velocity $V$ and GMC mass $M_{\rm GMC}$; reducing the impact parameter $b$ increases the energy gain, and thus the mass loss from the cluster, more quickly than it increases the angular momentum gain. In principle, many weak encounters could spin-up a cluster via a random walk process; we show below that this is infeasible given our simulation results.  

From Figure~\ref{fig: vrot}, we see that the difference in rotational velocity for a single encounter has a spread of order $ \Delta v_{\rm rot} = 0.01 ~\unit{km ~s^{-1}}$. Considering a random walk of encounters with the encounter parameters tested in our simulations, it would take $N \approx 10^4$ encounters to spin a cluster up from zero rotational velocity from $v_{\rm rot} = 1 ~\unit{km ~s^{-1}}$, taken from $v_{\rm rot} = \sqrt{N} \Delta v_{\rm rot}$. We can compare this to the expected number of interactions a cluster orbiting in the Solar neighborhood with Galactocentric radius $R_{\rm gal} \approx 8 ~\unit{kpc}$ will undergo with GMCs of mass $M_{\rm GMC} > 10^4 \unit{M_{\odot}}$, $b < 100 ~\unit{pc}$, and $V = 20 ~\unit{km ~s^{-1}}$. We estimate the number density of GMCs with masses $> 10^4 ~\unit{M_{\odot}}$ using the fitted surface density of GMCs as a function of Galactocentric radius given by \cite{Miville-Deschenes_2017}, i.e., $\Sigma = 83 \exp(-R_{\rm gal} / 2 ~\unit{Kpc}) \unit{~M_{\odot} ~pc^{-2}}$, and convert this to a number density by taking $n = \Sigma / 2 \langle M_{\rm GMC}\rangle H$, where $ \langle M_{\rm GMC}\rangle$ is the average mass of GMCs more massive than $10^4 ~\unit{M_{\odot}}$, and $H$ is the characteristic scale height of the disk. We assume $\langle M_{\rm GMC}\rangle = 8.4 \times 10^5 ~\unit{M_{\odot}}$ \citep{Miville-Deschenes_2017} and $H = 50 ~\unit{pc}$ \citep{Jeffreson_2022, Fujimoto_2023}. This corresponds to a number density of $n \approx 2 \times 10^{-8} ~\unit{pc^{-3}}$. The encounter rate is then computed as $\Gamma = n \sigma V = n \pi b_{\rm max}^2 V \approx 1.2 \times 10^{-2} ~\unit{Myr^{-1}}$, where $b_{\rm max} = 100 ~\unit{pc}$ and $V = 20 ~\unit{km ~s^{-1}}$. Over $1 ~\unit{Gyr}$, this encounter rate corresponds to $< 20$ such encounters. Even when ignoring the two-body relaxation processes that would ``spin-down" the cluster after each successive encounter, we conclude that a long-term ``spin-up" of OCs in the Solar neighborhood via a random walk of GMC--cluster encounters is not feasible; this matches the population statistics of rotating OCs, where rotators older than $1 ~\unit{Gyr}$ are uncommon \citep[e.g.,][]{Guilherme-Garcia_2023, Jadhav_2024, Dalessandro_2026}. Note that this order of magnitude calculation applies only to spherical clusters; clusters with significant rotation may be flattened \citep[e.g.,][]{Bianchini_2013, Kamann_2018}, for which $\Delta v_{\rm rot}$ may be significantly larger than $\sim 0.01 ~\unit{km ~s^{-1}}$, though testing this effect is outside of the scope of this work. 

In this work, we make no attempt to explain the population of rotating OCs, as the population is mostly consistent with the ``spin-down" of primordial rotation via coupling of two-body relaxation effects with the Galactic tides \citep{Dalessandro_2026}. We do, however, seek to provide insight into the few anomalously rotating dynamically old OC detections that do exist, such as M67. \cite{Sharma_2025} report that M67 has a rotational velocity of $0.81 \pm 0.07 ~\unit{km ~s^{-1}}$. This value is well above the predicted value of rotation that a fully relaxed cluster should sit at due to the Galactic tides of $\approx 0.1 ~\unit{km ~s^{-1}}$, following arguments by \cite{Tiongco_2016, Tiongco_2018}. This rotational velocity is also far too large to have come from primordial origin alone. \cite{Bianchini_2026} find that the peak of the rotation curve for a cluster at some epoch relative to its initial rotation is strongly related to the mass loss history of the cluster. Had M67 lost roughly $90\%$ of its initial mass \citep[as predicted by the evolutionary models of][]{Hurley_2005}, its initial $V_{\rm peak}$ would have needed to be $> 100 ~\unit{km ~s^{-1}}$ to correspond to its current $V_{\rm peak} \approx 1 ~\unit{km ~s^{-1}}$, which is non-physical. Other external dynamical effects that could affect the cluster's angular momentum, like disk shocking or a binary cluster merger, both seem implausible, as disk shocks are efficient in isotropizing orbits \citep{Oh_1992}, and any merger event would have needed to be within the first $\sim 100 ~\unit{Myr}$ of M67's lifetime \citep{Priyatikanto_2016}. 

While our simulations show that inducing rotation in clusters via the integrated effect of GMC flyby encounters is infeasible, a cluster that recently underwent a sufficiently energetic interaction can appear as if it has net rotation within a predefined tidal radius. We can define a post-encounter timescale of $\Delta t = 5 ~\unit{Myr}$, and compare this with the timescale for a disruptive GMC--cluster encounter $t_{\rm dis}^{\rm single}$, using eq.~28 in \cite{Gieles_2006}. We estimate $t_{\rm dis}^{\rm single} \approx 900 ~\unit{Myr}$ for a cluster of mass $500 ~\unit{M_{\odot}}$ and half-mass radius $3 ~\unit{pc}$, using a GMC mass density of $\rho = n \langle M_{\rm GMC} \rangle \approx 1.5 \times 10^{-2} ~\unit{M_{\odot} ~pc^{-3}}$, where all other values in eq. 28 are described by \cite{Gieles_2006}. The fraction of open clusters that have undergone such a destructive interaction within the past $5 ~\unit{Myr}$ is then $\Delta t / t_{\rm dis}^{\rm single} \approx 0.006$. Comparing this fraction with the $\approx 3500$ observed OCs within the catalog from \cite{Hunt_2024}, we find that $\approx 20$ of these observed clusters have had a disruptive GMC encounter within the last $5 ~\unit{Myr}$; a small, but non-negligible amount. We estimate that roughly $7$, $8$, and $84$ such recently disrupted clusters exist within the OC catalogs produced by \cite{Cantat-Gaudin_2018}, \cite{Castro-Ginard_2022}, and \cite{Perren_2023}, respectively. \cite{Hunt_2024} also derive an estimated number of OCs within the Milky Way as a whole of $1.3 \times 10^5$; scaling $\Delta t / t_{\rm dis}^{\rm single}$ with this value gives $\approx 750$ very recently disrupted OCs in the Milky Way. We emphasize that these estimates are order of magnitude only, as the GMC density depends strongly on Galactocentric radius \citep[e.g.,][]{Miville-Deschenes_2017, Kokaia_2019} and position relative to the Galactic spiral arms \citep[e.g.,][]{Stark_2006, Roman-Duval_2010}. 

While M67 sits $\approx 400 ~\unit{pc}$ above the disk \citep{Friel_1995, VandenBerg_2004}, making a very recent GMC--cluster encounter unlikely, there is tentative evidence for some dynamical interaction. M67 likely has tidal tails \citep{Gao_2020, Koz_2024, Sharma_2025}, and there may be misalignment with the principle component of M67's tidal tails and its orbital velocity vector \citep[see Figure 2 in][]{Sharma_2025}. \cite{Dinnbier_2022} show that old clusters on circular orbits should have tidal tails well aligned with their orbital velocity vector, and \cite{Jorgensen_2026} show that GMC--cluster encounters can produce scatter in the relative inclination angle of a cluster's tidal tails. We make no definitive claim of such an interaction, and only posit its plausibility as an explanation for detections of anomalously rotating, dynamically old OCs, similar to M67. 

The contents of \gaia DR4, scheduled for release in December 2026, will provide extraordinary insight into the population of rotating OCs. Characterizing bulk rotation at a significant level (i.e., $> 1 \sigma$) in OCs is a challenging endeavor; one needs an accurate membership catalog, precise astrometric/radial velocity information, and enough member stars to provide a statistically significant sample. Only $\approx 26\%$ of the rotating OCs reported by \cite{Dalessandro_2026} are found to rotate at the $1\sigma$ level, for example. The precisions of parallaxes, proper motions, and radial velocities in \gaia DR4 will improve by a factor of $\sqrt{2}$, $2 \sqrt{2}$, and $\sqrt{2}$, respectively, when compared with \gaia DR3 \citep{Brown_2025}. Moreover, the number of sources with RVS radial velocities will increase from $33$ million to over $100$ million \citep{Katz_2023}, allowing for more clusters to be studied in three velocity dimensions \citep[e.g.,][]{Sollima_2019, Dalessandro_2024}. These aforementioned data quality improvements will help to resolve differences between existing catalogs of rotating OCs \citep[e.g.,][]{Kuhn_2019, Guilherme-Garcia_2023, Jadhav_2024, Dalessandro_2026}, confirm or deny rotation in candidates with relatively large uncertainties, eliminate non-rotators with misattributed rotating signatures (e.g., apparent rotation due to the non-isotropic velocities of extra-tidal members), discover new rotation candidates, and shed light on the internal kinematics and dynamical histories of old rotation candidates like M67 and Ruprecht 147.

We note that the model and simulations presented in this work do not come without limitations. Our model for the approximate change in angular momentum due to a GMC--cluster encounter applies only for initially spherical clusters, and the internal energy gain of the cluster $\Delta E$ shows larger variations when compared with the analytic expectation from \cite{Gieles_2006} at larger impact parameters. However, these regimes (small $M_{\rm GMC}$, large $b$) coincide with encounters that should produce more negligible changes to the cluster's angular momentum, while both our model and simulations perform well at higher $M_{\rm GMC}$ and lower $b$. We make several simplifying assumptions in our simulations that we do not expect to alter the qualitative aspects of our results; we ignore primordial binaries, stellar evolution, Galactic tides, and the size of the GMC. Nevertheless, we do not expect any of these assumptions to change our conclusions in this work. We also emphasize that while this work is tailored towards encounters between GMCs and OCs, our model is directly applicable to encounters between dark matter subhaloes and GCs.

%% file: conclusions.tex
Inspired particularly by the large rotational velocity of the dynamically old OC M67, along with the pending release of \gaia DR4, we explore in this work the evolution of angular momentum, and thus rotation, of a simulated OC that underwent a nearby encounter with a point mass GMC. We derive an analytic expectation for the change in angular momentum induced by an encounter between a GMC and a star cluster, verify this expectation numerically using the $N$-body codes \texttt{REBOUND} and \texttt{REBOUNDx}, and explore the post-encounter rotational properties of the remaining subset of bound stars. Our main findings can be summarized as follows:

\begin{itemize}[label=--]
    
    \item The change in angular momentum of all stars in a cluster of mass $M_c = 1000 ~\unit{M_{\odot}}$ and $r_h = 3 ~\unit{pc}$ due to an impulsive GMC--cluster encounter, whether or not the stars remain bound to the cluster, is well described by Equation~\ref{eq: delta_Lx_approx} in the tested regime of encounter parameters, i.e., impact parameter at infinity $b = 5, 10, 20, 40, 60, 80, 100 ~\unit{pc}$, relative velocity at infinity $V = 20 ~\unit{km ~s^{-1}}$, and GMC mass $M_{\rm GMC} = 10^4, 10^5 ~\unit{M_{\odot}}$. 

    \item When allowing ample time for stars unbound by the encounter to escape the system, the rotational velocity of the bound subset of stars is unchanged when compared with its original pre-encounter value. Any changes to the rotational velocity of the bound subset post-encounter is consistent with the control simulations, where stars ejected from the system via two-body encounters may carry away angular momentum. 

    \item Because these encounters produce only very small changes to the rotational velocity in the bound subset of stars post-encounters, we rule out the possibility of a long-term ``spin-up" of a spherical cluster via weak perturbations from passing GMCs for clusters in the Solar neighborhood, as the necessary number of encounters to produce such a ``spin-up" is orders of magnitude larger than the expected number of GMC--cluster encounters given the local number density of GMCs.

    \item When determining boundedness purely via a radial criterion, as opposed to an energy one, a cluster that has undergone a very recent encounter can \textit{appear} as if it has significant rotation, as stars that have not yet had time to leave the system on the cluster outskirts can carry net-rotation. 

    \item A small, but non-negligible number of OCs ($\approx 0.6\%$) have had a very recent disruptive encounter with a GMC. We estimate that roughly $20$ of these exist within the OC catalog produced by \cite{Hunt_2024}, and roughly $750$ of these very recently disrupted systems exist within the Milky Way at large. Such systems, in the throes of their disruption, may briefly be confused as rotating bound clusters.

    \item The rotation reported in M67 by \cite{Sharma_2025} is truly anomalous; a primordial origin, tidally induced rotation, disk-shocking, and a binary cluster merger, all are implausible. We offer a recent perturbative encounter with a GMC as a potential explanation for M67's apparent rotation, but make no definitive claims. \gaia DR4 will help to confirm or deny the existence of rotation within M67's tidal boundary. 
\end{itemize}

The rotational properties of open star clusters are still relatively understudied. Only very recently have efforts to simulate \citep[e.g.,][]{Jerabkova_2021, Bissekenov_2025, Dalessandro_2026} and identify \citep[e.g.,][]{Kuhn_2019, Hao_2022, Guilherme-Garcia_2023, Hao_2024, Jadhav_2024, Dalessandro_2026} rotating OCs been undertaken. To our knowledge, we present the first detailed study on an origin of cluster rotation that is not primordial. At the advent of the release of \gaia DR4, studies like this one are necessary to provide insights into the underlying dynamical properties and histories of the soon-to-be-growing population of rotating open star clusters.

%% file: appendix.tex
Consider a star with mass $m_i$ in a star cluster whose center of mass lies at the origin, undergoing an encounter with a GMC of mass $M_{\rm GMC}$. We work in the frame of the cluster center of mass, and write the star's position as $\mathbf{r}_i(t) = \mathbf{r}_i + \delta \mathbf{r}_i(t)$, where $\mathbf{r}_i$ is its position before the encounter and $\delta \mathbf{r}_i(t)$ is its displacement relative to the cluster center of mass due purely to the GMC passage. In the impulse approximation, $\mathbf{r}_i$ may be treated as a constant. The acceleration of the star may then be written as $\mathbf{a}_i (t) = \ddot{\delta \mathbf{r}}_i (t)$, the explicit form of which we give in Equation~\ref{eq: a_rel}. The torque this star experiences is $\boldsymbol{\tau}_i (t) =  \mathbf{r}_i (t) \times m_i \ddot{\delta \mathbf{r}}_i (t)$, such that its change in angular momentum is $\Delta \mathbf{L}_i = m_i\int_{-\infty}^{\infty} \mathbf{r}_i (t) \times  \ddot{\delta \mathbf{r}}_i (t) \, dt$. Expanding this integral gives
\begin{equation}\label{eq: delta_Li_split}
    \Delta \mathbf{L}_i = m_i\, \mathbf{r}_i \times \int_{-\infty}^{\infty} \ddot{\delta \mathbf{r}}_i(t)\, dt \; + \; m_i \int_{-\infty}^{\infty} \delta \mathbf{r}_i (t)\times \ddot{\delta \mathbf{r}}_i(t)\, dt.
\end{equation}
The first integral in Equation~\ref{eq: delta_Li_split} is simply the final velocity kick due to the GMC passage, which may be written as
\begin{equation}\label{eq: final_vel}
    \int_{-\infty}^{\infty} \ddot{\delta \mathbf{r}}_i(t)\, dt = \Big[ \dot{\delta \mathbf{r}}_i (t) \Big]_{-\infty}^{\infty} \equiv \Delta \mathbf{v}_i,
\end{equation}
where $\lim_{t\rightarrow -\infty} \dot{\delta\mathbf{r}}_i (t) = 0$.  The second integral may be evaluated by parts as
\begin{equation}\label{eq: by_parts}
\begin{aligned}
    \int_{-\infty}^{\infty} \delta \mathbf{r}_i(t) \times \ddot{\delta \mathbf{r}}_i(t)\, dt &= \Big[ \delta \mathbf{r}_i(t) \times \dot{\delta \mathbf{r}}_i(t) \Big]_{-\infty}^{\infty} \\
    &\quad - \int_{-\infty}^{\infty} \dot{\delta \mathbf{r}}_i(t) \times \dot{\delta \mathbf{r}}_i(t)\, dt,
\end{aligned}
\end{equation}
where the second integrand in Equation~\ref{eq: by_parts} is zero. At the lower limit, both $\delta\mathbf{r}_i (t)$ and $\dot{\delta\mathbf{r}}_i (t)$ vanish. The upper limit may be written by noting that, at late times, the star drifts freely along its final velocity kick, such that $\delta \mathbf{r}_i(t) \rightarrow \Delta \mathbf{r}_i + \Delta \mathbf{v}_i\, t$, where we define the displacement of the star due to the encounter as
\begin{equation}\label{eq: delta_r_def}
    \Delta \mathbf{r}_i \equiv \lim_{t \rightarrow \infty} \left[ \delta \mathbf{r}_i(t) - \Delta \mathbf{v}_i\, t \right].
\end{equation}
The upper boundary of the first term in Equation~\ref{eq: by_parts} may then be written as
\begin{equation}\label{eq: upper_limit}
\begin{aligned}
    \lim_{t \rightarrow \infty} \delta \mathbf{r}_i(t) \times \dot{\delta \mathbf{r}}_i(t) &= \Delta \mathbf{r}_i \times \Delta \mathbf{v}_i + \lim_{t \rightarrow \infty} t\, \Delta \mathbf{v}_i \times \Delta \mathbf{v}_i \\
    &= \Delta \mathbf{r}_i \times \Delta \mathbf{v}_i,
\end{aligned}
\end{equation}
since $\lim_{t \rightarrow \infty} \dot{\delta \mathbf{r}}_i (t) = \Delta \mathbf{v}_i$. Equation~\ref{eq: delta_Li_split} then simplifies to
\begin{equation}\label{eq: delta_Li}
    \Delta \mathbf{L}_i = m_i \left( \mathbf{r}_i + \Delta \mathbf{r}_i \right) \times \Delta \mathbf{v}_i.
\end{equation}
To define the angular momentum change of the entire star cluster, we simply sum over all $i$ stars, such that
\begin{equation}\label{eq: delta_L_def_app}
    \Delta \mathbf{L} = \sum_i \Delta \mathbf{L}_i = \sum_i m_i \left( \mathbf{r}_i + \Delta \mathbf{r}_i \right) \times \Delta \mathbf{v}_i,
\end{equation}
where, since $\mathbf{a}_i (t)$ is measured relative to the accelerating cluster center of mass, each star also experiences a torque from the uniform acceleration of the center of mass; this vanishes in the sum since $\sum_i m_i \mathbf{r}_i = 0$ and, to the order considered here, $\sum_i m_i\, \delta \mathbf{r}_i = 0$. For simplicity, we write $\Delta \mathbf{L} = \Delta \mathbf{L}^{(v)} + \Delta \mathbf{L}^{(r)}$, where $\Delta \mathbf{L}^{(v)} \equiv \sum_i m_i \mathbf{r}_i \times \Delta \mathbf{v}_i$ and $\Delta \mathbf{L}^{(r)} \equiv \sum_i m_i \Delta \mathbf{r}_i \times \Delta \mathbf{v}_i$, and derive each term individually.

To derive $\Delta \mathbf{L}^{(v)}$, we note that \cite{Binney_Tremaine_2008} derive the final velocity kick of some mass $M$, which we set equal to $M_{\rm GMC}$, due to a hyperbolic encounter with a star of mass $m_i$, positioned at $\mathbf{r}_i = (x_i, y_i, z_i)$, in the center of mass frame, as
\begin{equation}\label{eq: vel_kick_bt}
\begin{aligned}
    \Delta \mathbf{v}_{M_{\rm GMC}^{\perp}} &= \frac{2 m_i V}{m_i + M_{\rm GMC}} \frac{b_i/b_{90}}{1 + (b_i/b_{90})^2} \left(- \hat{\mathbf{b}}_i\right), \\
    \Delta \mathbf{v}_{M_{\rm GMC}^{\parallel}} &= \frac{2 m_i V}{m_i + M_{\rm GMC}} \frac{1}{1 + (b_i/b_{90})^2} \left(- \hat{\mathbf{V}}\right),
\end{aligned}
\end{equation}
where $V$ is the relative velocity at infinity, $b_i$ is the impact parameter with respect to the $i$-th, $b_{90} \equiv G M_{\rm GMC} / V^2$, and $\hat{\mathbf{V}}$ and $\hat{\mathbf{b}}_i$ are the directions of the relative velocity and impact parameter at infinity, respectively. Note that $b_{90}$ in its exact form features the sum of the two masses involved in the encounter, but we consider the GMC to be sufficiently more massive than any individual stellar mass $m_i$ or the cluster mass $M_c$, such that $b_{90} = G M_{\rm GMC} / V^2$ is a reasonable approximation. In this work, we let $\mathbf{V} = (0, 0, V)$ and $\mathbf{b}_i = (-x_i, b-y_i, 0)$. We approximate the GMC's trajectory as a straight line, unperturbed by the cluster's gravity. Since we ignore any motion of the star due to the cluster potential, $M_{\rm GMC} \Delta \mathbf{v}_{M_{\rm GMC}} + m_i \Delta \tilde{\mathbf{v}}_i = 0$. Approximating $m_i + M_{\rm GMC} \approx M_{\rm GMC}$, we may write the final velocity kick of the star $\Delta \tilde{\mathbf{v}}_i$ in Cartesian coordinates as 
\begin{equation}\label{eq: delta_v_star}
\Delta \tilde{\mathbf{v}}_i = \frac{2V}{1 + (b_i/b_{90})^2} \left( \frac{-x_i}{b_{90}}, \frac{b - y_i}{b_{90}}, 1 \right).
\end{equation}
Defining the center of mass to be a particle of mass $M_c$, centered at the origin, we can similarly write the velocity kick of the center of mass to be 
\begin{equation}\label{eq: delta_v_com}
\Delta \tilde{\mathbf{v}}_{\rm com} = \frac{2V}{1 + (b/b_{90})^2} \left(0, \frac{b}{b_{90}}, 1 \right).
\end{equation}
We may write the final velocity kick of the $i$-th star relative to the cluster center of mass, i.e., $\Delta \mathbf{v}_i \equiv \Delta \tilde{\mathbf{v}}_i -\Delta \tilde{\mathbf{v}}_{\rm com}$, as
\begin{equation}
\label{eq: delta_v_star_rel}
\begin{aligned}
    \Delta \mathbf{v}_i = 2V\Bigg(&
    \frac{-x_i}{b_{90}}
    \frac{1}{1+(b_i/b_{90})^2},\\
    &
    \frac{b-y_i}{b_{90}}
    \frac{1}{1+(b_i/b_{90})^2}
    -\frac{b}{b_{90}}\frac{1}{1+(b/b_{90})^2},\\
    &
    \frac{1}{1+(b_i/b_{90})^2}
    -\frac{1}{1+(b/b_{90})^2}
    \Bigg).
\end{aligned}
\end{equation}

Assuming the cluster follows a spherically symmetric Plummer density profile, all but one term in $\Delta \mathbf{L}^{(v)}$ vanishes under summation; the $x$-component of $\Delta \mathbf{L}^{(v)}$ is $\Delta L_x^{(v)} = \sum_i m_i (y_i \Delta v_{z,i} - z_i \Delta v_{y,i})$, where the first term may be non-zero since $y_i \Delta v_{z,i}$ is even in $x_i$ and $z_i$ and has no definite parity in $y_i$ due to the GMC's passage at $y=b$. $\Delta \mathbf{L}^{(v)}$ can be written as
\begin{equation}\label{eq: delta_Lv_cluster_comp}
    \Delta \mathbf{L}^{(v)} = 2 V \left( \sum_i m_i \frac{y_i}{1 + (b_i/b_{90})^2}, 0, 0 \right),
\end{equation}
where the center of mass term vanishes since $\sum_i m_i y_i = 0$. For a large number of particles $N$, we may treat the sum in Equation~\ref{eq: delta_Lv_cluster_comp} as an integral, such that
\begin{equation}\label{eq: delta_Lv_cluster_int}
    \Delta L_x^{(v)} \approx 2 V \int  \frac{y}{1+(b_{\rm eff}/b_{90})^2} \rho(r) d^3r,
\end{equation}
where $b_{\rm eff} = \sqrt{x^2 + (b - y)^2}$, $\rho(r)$ is the Plummer density profile $\rho(r) = (3 M_c / 4\pi a^3 ) \left[ 1 + r^2 / a^2 \right]^{-5/2}$, and $a$ is the Plummer radius, related to the half-mass radius by $r_h = 1.305 a$. Note that in the weak-encounter limit ($b_{\rm eff} \gg b_{90}$), Equation~\ref{eq: delta_Lv_cluster_int} is $\mathcal{O}(b_{90}^2/b_{\rm eff}^2)$, since $[1 + (b_{\rm eff}/b_{90})^2]^{-1} \simeq b_{90}^2 / b_{\rm eff}^2$.

Equation~\ref{eq: delta_v_star_rel} gives only the velocity kick of the $i$-th star long after the encounter ($t \rightarrow \infty$), so to obtain $\Delta \mathbf{r}_i$ (and thus, $\Delta \mathbf{L}^{(r)}$), we instead follow the star's response to the GMC over the duration of the encounter. The acceleration of the $i$-th star relative to the cluster center of mass, introduced above, is
\begin{equation}\label{eq: a_rel}
    \mathbf{a}_i (t) = \frac{GM_{\rm GMC}}{|\mathbf{X}(t) - \mathbf{r}_i|^3} \left( \mathbf{X}(t) - \mathbf{r}_i \right) - \frac{GM_{\rm GMC}}{|\mathbf{X}(t)|^3} \mathbf{X}(t),
\end{equation}
where $\mathbf{X}(t) = (0, b, Vt)$ is the position of the GMC relative to the origin at time $t$, such that the closest approach of the GMC to the cluster center is at $t=0$. While integrating Equation~\ref{eq: a_rel}, we hold $\mathbf{r}_i$ fixed, neglecting the effect of the star's displacement due to the GMC passage on the acceleration it experiences. Writing the star's position during the encounter as $\mathbf{r}_i + \delta \mathbf{r}_i(t)$ and Taylor expanding Equation~\ref{eq: a_rel} about $\mathbf{r}_i$, the relative acceleration of the $i$-th star may be written as
\begin{equation}\label{eq: a_taylor}
    \mathbf{a}_i(\mathbf{r}_i + \delta \mathbf{r}_i(t), t) = \mathbf{a}_i(\mathbf{r}_i, t) + \left( \delta \mathbf{r}_i \cdot \nabla \right) \mathbf{a}_i \big|_{\mathbf{r}_i} + \mathcal{O}\!\left( |\delta \mathbf{r}_i|^2 \right).
\end{equation}
Since $\mathbf{a}_i (t) \propto M_{\rm GMC}$ and $\delta \mathbf{r}_i (t)$ is itself $\mathcal{O}(b_{90}/b_{\rm eff})$, the second term is $\mathcal{O}(b_{90}^2/b_{\rm eff}^2)$. Holding $\mathbf{r}_i$ fixed retains only the first term, such that the neglected terms contribute to $\Delta \mathbf{L}^{(r)}$ only at $\mathcal{O}(b_{90}^3/b_{\rm eff}^3)$, negligible compared with the $\mathcal{O}(b_{90}^2/b_{\rm eff}^2)$ terms retained in $\Delta \mathbf{L}$. We note that the fixed-$\mathbf{r}_i$ treatment is only used to obtain $\Delta \mathbf{r}_i$. For $\Delta \mathbf{L}^{(v)}$ we use the exact final two-body velocity kick (Equation~\ref{eq: delta_v_star_rel}), since $\Delta v_{z,i}$, which alone contributes to Equation~\ref{eq: delta_Lv_cluster_int}, first appears at $\mathcal{O}(b_{90}^2/b_{\rm eff}^2)$. In $\Delta \mathbf{L}^{(r)}$, by contrast, only the $\mathcal{O}(b_{90}/b_{\rm eff})$ part of $\Delta \mathbf{v}_i$ contributes since $\Delta \mathbf{r}_i$ is also $\mathcal{O}(b_{90}/b_{\rm eff})$, for which Equation~\ref{eq: delta_v_star_rel} and Equation~\ref{eq: delta_v_time} agree.

Holding $\mathbf{r}_i$ fixed, we may write the separation between the GMC and $i$-th star as $\mathbf{X}(t) - \mathbf{r}_i = (-x_i, b-y_i, Vt - z_i)$ with magnitude $R_{i}(t) \equiv \left[ b_i^2 + (Vt - z_i)^2 \right]^{1/2}$, and similarly $R(t) \equiv (b^2 + V^2 t^2)^{1/2}$. Using the substitution $u = Vt'-z_i$,  integrating Equation~\ref{eq: a_rel} with $\mathbf{r}_i$ fixed from $t' = - \infty$ to $t' = t$ yields the $i$-th star's velocity kick at time $t$, $\Delta \mathbf{v}_i(t) \equiv \dot{\delta \mathbf{r}}_i(t)$, given by
\begin{equation}\label{eq: delta_v_time}
\begin{aligned}
    \Delta \mathbf{v}_i (t) = V b_{90} \Bigg(& - \frac{x_i}{b_i^2} \left[ 1 + \frac{V t - z_i}{R_i} \right], \\
    & \frac{b - y_i}{b_i^2} \left[ 1 + \frac{V t - z_i}{R_i} \right] \\
    & \quad - \frac{1}{b} \left[ 1 + \frac{V t}{R} \right], \\
    & - \frac{1}{R_i} + \frac{1}{R} \Bigg),
\end{aligned}
\end{equation}
where the components follow from the anti-derivatives of $(b_i^2 + u^2)^{-3/2}$ and $u\,(b_i^2 + u^2)^{-3/2}$, giving the perpendicular ($x, y$) and parallel ($z$) components of Equation~\ref{eq: delta_v_time}, respectively. The center of mass terms follow identically with $z_i \rightarrow 0$ and $b_i \rightarrow b$. Note that expanding to first order in $b_{90}/b_i$ and $b_{90}/b$ in Equation~\ref{eq: delta_v_star_rel} is equivalent to Equation~\ref{eq: delta_v_time} under the limit of $t \rightarrow \infty$. 

Under our definition of $\Delta \mathbf{r}_i$ in Equation~\ref{eq: delta_r_def}, we integrate Equation~\ref{eq: delta_v_time} from $t' = -\infty$ to $t' = t$, subtract the free drift $\Delta \mathbf{v}_i\, t$, and take $t \rightarrow \infty$, giving
\begin{equation}\label{eq: delta_r}
    \Delta \mathbf{r}_i = 2 b_{90} \left(\frac{x_i z_i }{b_i^2}, - \frac{(b - y_i)z_i}{b_i^2}, \ln{\frac{b_i}{b}} \right), 
\end{equation}
where we again use the substitution $u = Vt' - z_i$ and utilize the anti-derivatives of $1 + u\,(b_i^2 + u^2)^{-1/2}$ for the $x$ and $y$ components of $\Delta \mathbf{r}_i$, and $(b_i^2 + u^2)^{-1/2}$ for the $z$ component. 

The spherical symmetry of the system again causes all but one term in $\Delta \mathbf{L}^{(r)}$ to cancel due to odd parity in either $x_i$ or $z_i$. Specifically, the second term in $\Delta L_x^{(r)} = \sum_i m_i (\Delta y_i \Delta v_{z,i} - \Delta z_i \Delta v_{y,i})$ is non-zero, as this term is even in $x_i$ and $z_i$ and has no definite parity in $y_i$. We may write $\Delta \mathbf{L}^{(r)}$ as
\begin{equation}\label{eq: delta_Lr_cluster_comp}
    \Delta \mathbf{L}^{(r)} = 2 b_{90} \left( -\sum_i m_i \ln{\left( \frac{b_i}{b} \right) } \Delta v_{y,i} (x_i, y_i), 0, 0 \right),
\end{equation}
where $\Delta v_{y,i} (x_i, y_i)$ is the $y$ component of Equation~\ref{eq: delta_v_star_rel}. Approximating Equation~\ref{eq: delta_Lr_cluster_comp} as an integral over the density profile, and combining with Equation~\ref{eq: delta_Lv_cluster_int}, we arrive at Equation~\ref{eq: delta_Lx_int}, and write the change in internal angular momentum of the star cluster in the $x$-direction as
\begin{equation}\label{eq: delta_Lx_int_app}
\begin{aligned}
    \Delta L_x^{\rm int} &= \Delta L_x^{(v)} + \Delta L_x^{(r)} \\
    &\approx 2V \int \frac{y}{1 + (b_{\rm eff}/b_{90})^2}\, \rho(r)\, d^3 r \\
    &\quad - 2 b_{90} \int \ln\!\left( \frac{b_{\rm eff}}{b} \right) \Delta v_y(x, y)\, \rho(r)\, d^3 r.
\end{aligned}
\end{equation}